\documentclass[sigconf]{acmart}
\usepackage{multirow}
\usepackage{tabularx}
\usepackage{enumitem}

\newcommand{\linguistic}[1]{{\textit{baseline+psycholinguistic}}}
\newcommand{\morality}[1]{{\textit{baseline+psycholinguistic+morality}}}

\acmConference[ESEC/FSE 2023]{The 31st ACM Joint European Software Engineering Conference and Symposium on the Foundations of Software Engineering}{11 - 17 November, 2023}{San Francisco, USA}

\AtBeginDocument{%
  \providecommand\BibTeX{{%
    \normalfont B\kern-0.5em{\scshape i\kern-0.25em b}\kern-0.8em\TeX}}}

\setcopyright{acmcopyright}
\copyrightyear{2023}
\acmYear{2023}
\acmDOI{XXXXXXX.XXXXXXX}

\acmPrice{15.00}
\acmISBN{978-1-4503-XXXX-X/18/06}

\begin{document}

\title{Exploring Moral Principles Exhibited in OSS: A Case Study on GitHub Heated Issues}

\author{Ramtin Ehsani}
\affiliation{%
  \institution{Drexel University}
  \city{Philadelphia, PA}
  \country{USA}}
\email{ramtin.ehsani@drexel.edu}

\author{Rezvaneh Rezapour}
\affiliation{%
  \institution{Drexel University}
  \city{Philadelphia, PA}
  \country{USA}}
\email{shadi.rezapour@drexel.edu}

\author{Preetha Chatterjee}
\affiliation{%
  \institution{Drexel University}
  \city{Philadelphia, PA}
  \country{USA}}
\email{preetha.chatterjee@drexel.edu}


\begin{abstract}
To foster collaboration and inclusivity in Open Source Software (OSS) projects, it is crucial to understand and detect patterns of toxic language that may drive contributors away, especially those from underrepresented communities. 
Although machine learning-based toxicity detection tools trained on domain-specific data have shown promise, their design lacks an understanding of the unique nature and triggers of toxicity in OSS discussions, highlighting the need for further investigation.
In this study, we employ Moral Foundations Theory to examine the relationship between moral principles and toxicity in OSS. Specifically, we analyze toxic communications in GitHub issue threads to identify and understand five types of moral principles exhibited in text, and explore their potential association with toxic behavior. Our preliminary findings suggest a possible link between moral principles and toxic comments in OSS communications, with each moral principle associated with at least one type of toxicity. The potential of MFT in toxicity detection warrants further investigation. 
\end{abstract}





\maketitle



\section{Introduction}


Open source software (OSS) projects have become crucial for the digital advancement of society, with over 90\% of companies leveraging open source~\cite{GithubOcto2022}. However, many new and longstanding OSS projects fail, in part due to toxic interactions and uncivil language, which serve as a barrier to contribution and lead to negative emotions and isolation~\cite{Igor15, qiu19, 9402044}. 
Thus, understanding toxicity in OSS has gained increasing attention in recent years~\cite{ cheriyanTowards2021, Sarker2020ABS, toxicr, ferreiraIncivility2022, ferreiraHeat2022, egelmanPredicting2020}. 
Uncivil features such as name-calling, frustration, and impatience were found in more than half of non-technical emails in the Linux Kernel mailing list~\cite{ferreiraSTFU2021}, swearing and personal offense were identified as the main reasons for offense in several communities~\cite{cheriyanTowards2021}, and GitHub issue threads contain various types of toxic behaviors~\cite{Miller2022}. 
Although researchers outside the Software Engineering (SE) domain have designed several techniques to automatically detect toxicity in online forums~\cite{shethDefining2021, maityOpinion2018, pascualFerratoxicity2021}, directly applying those tools to SE is ineffective due to several factors, including the unique nature of SE text~\cite{Raman2020, Sarker2020ABS, Chatterjee2022DataAugmentation}. 
More recently, machine learning-based tools have shown promise in detecting toxicity when trained on domain-specific data~\cite{toxicr}, however it remains crucial to address the fundamental issue of their design lacking an understanding of the unique nature and triggers of toxicity in OSS~\cite{Miller2022,9793879}.

Language is an indispensable tool to convey information between people, and analyzing it could play a significant role in understanding the challenges faced in OSS communities ~\cite{saneiImpacts2021, ferreiraSentiment2019, Sajadi2023}. Linguistic styles and word choices can help in analyzing and understanding people's thoughts, opinions, and feelings~\cite{triandis1989self,boyd2022development,rezapour2019enhancing, Chatterjee21}. 
People's values and personal norms affect their (spontaneous) attitude, decision-making process, and what they perceive as good or bad, and moral or immoral \cite{haidt2001emotional,rezapour2021incorporating}. Previous research in psychology has addressed the connection between moral principles and hate and showed that morality is a key feature of hatred; hate is connected to core moral beliefs and higher levels of moral emotions (e.g., contempt, anger, disgust)~\cite{morality_hate_2018,morality_hate_2021}. 
Considering the importance of effective communication, we believe that analyzing moral values as exhibited in OSS texts can provide valuable insights into the underlying beliefs and values that influence communication styles in this community, especially in spreading toxicity. 

In this paper, we aim to investigate the moral principles exhibited in GitHub and explore their relationship with toxicity. 
To that end, we leverage Moral Foundation Theory (MFT), a social psychological theory that assumes individual judgments are influenced by emotional and cognitive appraisals, referred to as intuitions or foundations~\cite{graham2013moral,haidt2004intuitive}. 
We explore a dataset of 100 toxic issue threads labeled with their representative natures (e.g., `Insulting', `Arrogant')~\cite{Miller2022}, to analyze moral principles and how they are exhibited in the context of OSS project communications. We map the observed moralities in each thread to their corresponding natures of toxicity to understand the relationship and co-occurrence. 
Our findings reveal that toxic issue threads exhibit moral principles, with each moral principle being associated with at least one type of toxicity. This suggests that toxic comments are associated with moral principles, and that the Moral Foundations Theory (MFT) may be useful in detecting toxicity. 
To apply MFT to the domain of SE, it would be necessary to adapt it, and we believe that doing so has the potential to enhance our understanding and detection of toxicity in SE communications.

\vspace{-6pt}
\section{Moral Foundations Theory}
\label{moraldef}
With the goal of investigating instances on GitHub that represent moral principles exhibited in the text, we leverage Moral Foundations Theory (MFT)~\cite{graham2013moral}. MFT classifies human behavior into five fundamental principles that represent contrasting values: 

\noindent
\textbf{Care/harm}: This principle is based on our general dislike of suffering, whether for ourselves or others, related to our evolution as mammals. It underlies the virtues of kindness, compassion, and gentleness, and it condemns cruelty and aggression.

\noindent
\textbf{Fairness/cheating}: This principle  
relates to justice and rights, and is linked to the evolutionary process of reciprocal altruism.

\noindent
\textbf{Loyalty/betrayal}: This principle promotes patriotism, heroism, trust, and self-sacrifice for the group, rooted in our history as tribal beings. It values the virtue of "One for all, and all for one," while considering acts of betrayal towards social structures as immoral.

\noindent
\textbf{Authority/subversion}: This principle is shaped by our primate history of hierarchical social structure, resulting in virtues such as leadership, followership, and deference to authority/traditions. It may view dissent against authority as immoral.

\noindent
\textbf{Sanctity/degradation}: This principle arises from the psychology of disgust and contamination, where the concept of humans striving to live in a noble and elevated way is the key.  It reflects the 
idea of the body as a temple that can be corrupted by immorality.

\vspace{-6pt}
\section{Methodology} 
To understand how moral principles manifest in an OSS project's textual communication and its relation to toxicity, we analyze the GitHub developer interactions. 
Our focus is on GitHub heated issues, which often result from conflicts, as these issues can provide valuable insights into the ethical and moral considerations of participants.
It should be noted that we are only analyzing textual representations of moral values, and our analysis does not represent the moral values of 
people engaging in these discussions.


\newcolumntype{b}{>{\hsize=1.3\hsize}X}
\newcolumntype{s}{>{\hsize=.8\hsize}X}
\newcolumntype{k}{>{\hsize=.2\hsize}X}
\definecolor{myblue}{RGB}{0, 103, 148}

\begin{table*}[t]
\centering
\small
\resizebox{\textwidth}{!}{%
\begin{tabularx}{\textwidth}{|k|s|}
\hline
   \textbf{Moral Principle --- Def.} & \textbf{As Exhibited in Issue Threads}
  \\ \hline\hline
   \textcolor{myblue}{Care/harm} --- Protecting versus hurting others & \textbf{Care:} Developer communications exhibiting kindness, e.g., experienced contributors shielding newcomers from criticism. 
   
   \textbf{Harm:} Developers exchanging insults in issue threads, either in response to feature requests or bug fixes, or due to trolling or experience from past interactions~\cite{Miller2022}). 
   \\ \hline
   
   \textcolor{myblue}{Fairness/cheating} --- Cooperation/ trust/ just versus cheating in interaction with objects and people & \textbf{Fairness:} Developers and contributors respecting one another’s rights (e.g., having the right to open threads and ask for features/fixes, get the help they need). 
   
   \textbf{Cheating:} 
   Disregarding developers' rights by imposing unrealistic expectations (e.g., unfeasible project timeline), or by discriminating against them (e.g., based on gender) and not addressing their concerns.~\cite{vanbreukelen2023still}. 
\\ \hline
   \textcolor{myblue}{Loyalty/betrayal} --- Ingroup commitment (to coalitions, teams, brands) versus leaving group & \textbf{Loyalty:} Acts of self-sacrifice and altruism for the community, ranging from dedicating time to participating in issues threads to defending developers and the project itself against criticisms.
   
   \textbf{Betrayal:} OSS communities, like all social communities, view acts of betrayal as a breach of community morale. Betrayal is observed when developers urge others to abandon a project, or exclude them from the community. 
\\ \hline
   \textcolor{myblue}{Authority/subversion} --- Adhering to the rules of hierarchy versus challenging hierarchies & \textbf{Authority:} OSS communities often have guidelines that must be followed, e.g., Code of Conduct: Authorities enforcing CoC or any admin-privileged acts (minimizing comments, closing issues, etc.) to ban or censor the users. 
   
   \textbf{Subversion:} Developers trying to rebel against authority, and questioning contributors’ ability to lead the community.\\ 
   
   \hline
   \textcolor{myblue}{Sanctity/degradation} --- Behavioral immune system versus spontaneous reaction & \textbf{Sanctity:} Communicating with temperance and respect for one another, and encouraging others to do so as well. 
   
   \textbf{Degradation:} Developers expressing their hatred (disgust) toward a certain package, system, or code in the project. 
   This behavior is targeted at code (objects) rather than people. 
   \\
  \hline
\end{tabularx}
}
\caption{Moral Principles in GitHub Heated Issues}
\label{tab:orgDefs}
\vspace{-0.8cm}
\end{table*}


\noindent
\textbf{Dataset.}
We analyze a benchmark dataset of 100 toxic GitHub issue comment threads curated by Miller et al. \cite{Miller2022}, in which toxicity is defined as "rude, disrespectful, or unreasonable language that is likely to make someone leave a discussion". 
The dataset includes information on the issue link, author, trigger, target, and the nature of toxicity for each thread of comments. Out of the 100 issue threads, 20 contain comments that were either removed or the project along with its issues were taken down from GitHub.
The 5 categories of toxic instances in this dataset are: \textit{Insulting}, \textit{Entitled}, \textit{Arrogant}, \textit{Trolling}, and \textit{Unprofessional}.

\noindent
\textbf{Procedure.}
To understand the characteristics and patterns of morality in the context of SE, we qualitatively analyze 100 issue threads and their context,
using thematic analysis \cite{BraunThematicAnalysis}, and following the trustworthiness criteria (credibility, transferability, dependability, and confirmability) as recommended by previous research \cite{lincolnNaturalistic1985, nowellThematic2017}. 
The data analysis process was conducted iteratively and reflectively, following recommendations from qualitative analysis~\cite{qual}. 
To enhance our understanding of the developers' interactions and the contextual factors influencing their communication, we first carefully read through all the issue comment threads. Next, the goal was to associate moral values (as defined in Section \ref{moraldef}) to each comment in the threads, if applicable. Given the substantial differences between OSS communications and other social media platforms such as Twitter and Reddit, we adapted the original definitions of morality to align them with the context of software engineering.
Specifically, we identify key concepts of each morality type, observe patterns in the dataset, and use them to create the mapping of morality categories. 
Throughout the analysis, careful notes were taken to capture patterns of morality observed in the comments. This process was repeated multiple times, with constant additions to the notes and observations, and ongoing reevaluation of the mappings in each iteration.

Note that in the process of labeling morality to the issue comments, we consciously disregard the toxicity labels to mitigate bias in our decisions, so that we do not associate a specific type of toxicity with a certain type of morality.
The result of our mapping between the morality categories and their manifestation in GitHub heated issues is presented in Table \ref{tab:orgDefs}.
\vspace{-.2cm}
\section{Preliminary Observations}
\label{results}

In this section, we discuss our observations for the moral principles as exhibited in the GitHub heated issues. 
We analyze the issue comment threads in our data sample to identify instances where moral principles were expressed in the text.
Out of 695 issue comments across 100 threads, 135 exhibit at least one form of moral principle.
In this analysis, 
people who open the issue threads to find fixes for problems are referred to as \textit{users}, and people who try to resolve the issues and close the threads as \textit{contributors}. 
We observed unique patterns specific to each of the identified roles.


\noindent
\textbf{Care/Harm.}
A total of 48 comments exhibit this principle.
Pertaining to this category, following are the behaviors we observed: \textbf{(CH1) }users who use derogatory language or insults towards contributors while seeking assistance or requesting new features; \textbf{(CH2) }users insulting contributors mainly by trolling, possibly due to differences in ideology, opinions, etc.; \textbf{(CH3) }contributors responding using insults toward users.

The first type (CH1) was the most common in our dataset (32 out of 48 instances). For example, in one thread, a user opens an issue saying that the name of the project is duplicated and it should change while using insults: \textit{"This name is in use and should be changed immediately. Yes, I have seen the other issues. Yes, I am opening a new one because f**k you Microsoft. You are merely trying to cast a shadow on other, truly open source projects..."}.

The second type (CH2) was also seen in a number of cases (5 out of 48 instances). For example, in one instance, a user opens an issue saying: \textit{"Revenue. F**k you guys"}, and does not add anything else to the thread. Threads similar to this are mostly the result of trolling, but they can also be the result of differences in ideology or based on previous interactions~\cite{Miller2022}.

For the third type (CH3), we saw an interesting pattern of contributors responding in a detrimental manner towards the users. This type is rare compared to other types in this category.
In one instance, a user opens an issue asking: 
\textit{"Where to f**k python2? Why, when I give the brew install python command, python3 is installed, not python2. Developers are you stoned there?"},
and surprisingly, one of the contributors responds with: 
\textit{"I recommend not having intercourse with EOL software. Python 2 is no longer included in Homebrew"}. The user then responds: \textit{"You have an unfinished raw product with some problems. What can be your attitude? Treating users like sh*t"}. This issue thread exhibits the first and last patterns (CH1 and CH3).

\noindent
\textbf{Fairness/Cheating.}
We observed behaviors associated with this category in a total of 19 comments, which include: \textbf{(FC1) }users having unrealistic expectations from the contributors; \textbf{(FC2) }contributors failing to address user's issues due to unjustified reasons.

Among the instances of this category, 7 out of 19 instances exhibit the first type of communication pattern. In one of the instances representing the first type (FC1), the user has opened an issue asking about a problem in the debugger. Contributors of the projects try to help the user by providing tips on how the problem can be solved, but as the conversation goes on, the user gets frustrated and says: \textit{"@<USER1> @<USER2> Too busy to respond?"}, which in the end, leads to the issue thread being locked. This is an example of unrealistic expectations about  response time causing conflict.

For the second type (FC2), 
for instance, this is how one of the contributors responds to a  user's issue: \textit{"How would be this a priority, the app must be ready to be shipped in a few days and basically, no one uses the smartphone in landscape mode except for watching media"}.

Another type of behavior related to \textit{Fairness} would be discrimination (based on race, culture, gender, religion, etc). In our dataset, we did not observe any instances of discrimination. Analysis of additional data may reveal discriminatory behavior in OSS, as such occurrences have been observed in previous studies~\cite{vanbreukelen2023still, cheriyanTowards2021}.

\noindent
\textbf{Loyalty/Betrayal.}
The observed patterns for this category of morality were found in 23 instances of the dataset: \textbf{(LB1) }users actively promoting and inciting rebellion against a project, encouraging others to switch to an alternative project (most common with 15 instances); \textbf{(LB2:) }contributors excluding users from the project.

To better understand these behaviors, we provide examples for each type. For instance, in one of the issue threads, a user claims that SSL Insecure is not being respected in the project: \textit{"Whoever will find this issue and gets pissed off, because the author doesn't bother to fix it for years, ditch mitmproxy and use SSL SPLIT"}. One of the developers responds with: \textit{"I'm glad you have found something that works for you!"}. In this instance, both LB1 and LB2 were observed, because the user is encouraging people to leave this project behind, and the contributor also encourages the user to not use the project.
In another instance, a user requests the removal of the slur filter in the app, however, one of the contributors says: \textit{"If you don't like it, fork it. Stop bothering us about it, we will never fully remove the slur filter"}, which counts as excluding the user from the community.


\begin{figure*}[t] 
\centerline{\includegraphics[scale=0.18]{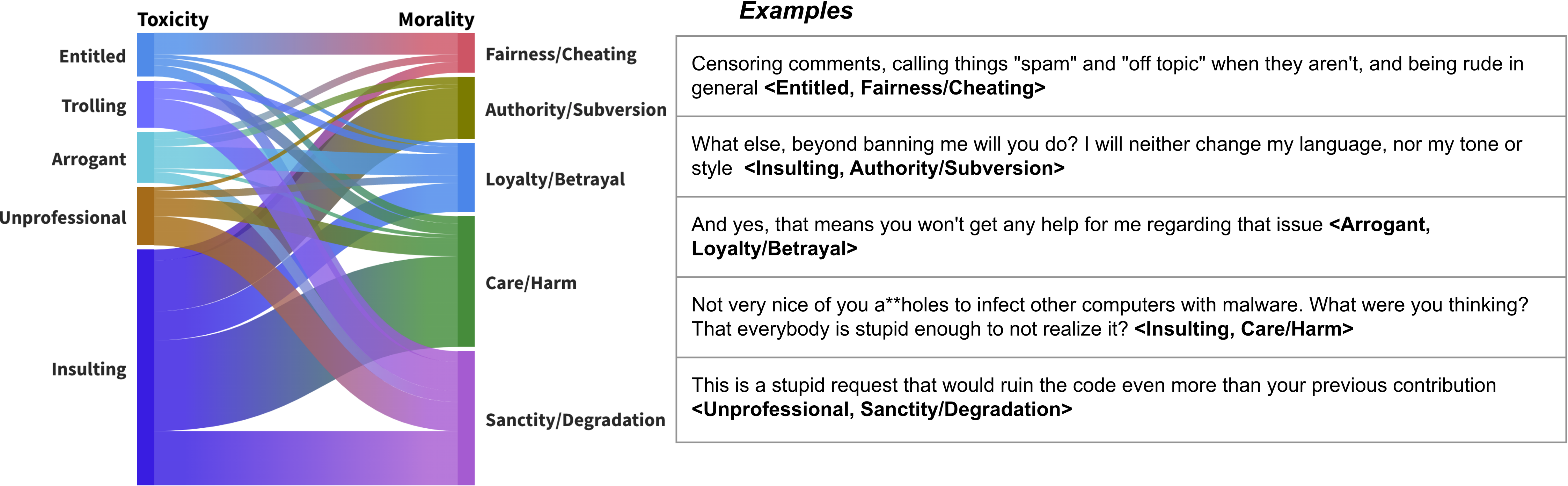}}
\vspace{-10pt}
\caption{Morality and Toxicity: Examples of their Co-occurrence as Exhibited in GitHub}
\vspace{-6pt}
\label{OSSMoralToxic}
\vspace{-0.25cm}
\end{figure*}

\noindent
\textbf{Authority/Subversion.}
OSS communities, like any other community, often have specific rules and guidelines that everyone has to follow. GitHub projects use the Code of Conduct (CoC) in an attempt to promote their expectations and standards of ethical behavior within the community \cite{touraniCode2017}. 
The establishment of such rules and hierarchies often leads to the dynamic interplay of moralities such as authority and subversion in OSS communications. A total of 30 instances with this principle were found in our dataset.
The patterns observed for this category are: \textbf{(AS1) }users trying to rebel against authority, and questioning developers' ability to lead the community. \textbf{(AS2) }contributors (authorities) enforcing Code of Conduct when thought necessary or any other admin-privileged acts to ban, censor, or silence users (most common with 16 instances).

In one of the instances, a user has opened an issue, claiming that the project has stolen his code: \textit{"Don’t be a d*ck as the source code was stolen at that point and implemented on this"}.
The contributor enforces authority by saying: \textit{"please be civilized and refrain from profanities as required by our Code of Conduct, or I'll ban you from all Falconry projects. This is the last warning"}.
And the user challenges the authority by saying: \textit{"I’m not a bot but if you wish to ignore me go ahead and I’ll get a lawyer"}.
Additionally, there were instances where the users questioned the capabilities of contributors. For example, a user opened an issue asking about how to do a certain thing within the system. As the conversation goes on, the user responds to one of the contributors: \textit{"...I would hate to have you in charge of any security issues"}.

\noindent
\textbf{Sanctity/Degradation.}
The most commonly observed pattern in this category, in a total of 42 instances, was observed when: (\textbf{SD1}) users express their hate (disgust) toward a certain package, system, or code that the project is using, most likely due to  their personal preferences. This principle is targeted at the code/system rather than people (targeting people represents \textit{Care/Harm}).

For example, a user opened an issue in the project expressing how much the document is poorly written, and in a sense, he is \textit{disgusted} with it: \textit{"...in fact, you write a sh*t doc. I'm a real man, it's my feeling of your holy sh*t doc"}. 
Or in another instance, a user opens an issue explaining his problem with the software: \textit{"...I just tried reinstalling you buggy, sh*tty software for the third time..."}.


After gathering our observations of the exhibited patterns of morality in GitHub communications, we associated the toxicities presented in our dataset with the moral principles in each issue thread. The results generated for this mapping are shown in Figure \ref{OSSMoralToxic}. Based on this mapping, we observe  that every type of toxicity is associated with at least one form of morality. 
Entitlement is mostly associated with the principle of \textit{Fairness/Cheating}, Trolling with \textit{Sanctity/Degradation}, Arrogant to \textit{Loyalty/Betrayal}, Unprofessional with \textit{Sanctity/Degradation}, and Insulting with \textit{Care/Harm}.

The relationship between the concepts of toxicity and moral principles can be observed in their respective definitions. For example, entitlement comes from disregarding people's rights for personal gain. This aligns with the \textit{Fairness/Cheating} definitions given in previous sections. 
The relationship between \textit{Authority/Subversion} and insulting threads implies that using insults causes the authorities to intervene more in the issues compared to the other types of toxicity.
The moralities most frequently observed in toxic threads are \textit{Sanctity/Degradation} and \textit{Care/Harm}, which were primarily expressed through insults directed towards code or individuals.

\vspace{-6pt}
\section{Implications and Challenges}
We found that toxic behaviors can be categorized and linked to various types of moral principles, indicating the potential for utilizing moral principles to detect toxicity in OSS communications. 
To this end, numerous Moral Foundations Dictionaries (MFD) can be integrated into models to detect moral values in SE texts, as demonstrated in other fields~\cite{rezapour2019enhancing,kennedyMoral2022}. 
Recent studies have explored the association between moral values and the sense of belonging in OSS communities~\cite{trinkenreich2023}.
Adapting existing morality dictionaries to SE and incorporating them into detection tools could facilitate a better understanding of several human values in OSS, including toxicity. 
It is worth noting that our study only examined the association between moral principles and toxicity in OSS communities, and did not explore any potential causal relationships between the two. While our findings suggest that moral principles are associated with toxic interactions, it is possible that other factors may also contribute to toxic behavior in OSS communities. 

We acknowledge that inferring human values based on textual communication data has its perils. For instance, previous studies show high error rate in detecting developer personalities based on textual data~\cite{vanmilPromises2021, calefatoUsing2022}. We emphasize that we are only analyzing the textual representation of moral principles, and we are not associating morality values with the people involved in the discussions.
Our findings of the exhibited moralities and their association with toxicity are limited to a relatively small dataset of GitHub issue threads. Further analysis on a large and diverse dataset is necessary to generalize the findings.  
Detecting morality even in social contexts (e.g., Twitter) is a hard task, mostly in distinguishing co-occurring moral principles~\cite{hooverMoral2020}. 
We attempted to address that challenge by mapping and creating clear definitions of moral principles in the domain of SE (Table \ref{tab:orgDefs}).
Another important point to consider is the potential impact of cultural and social differences on the interpretation and expression of moral principles in OSS communities. Our study focused on English texts, and it is possible that different cultures and languages may have different conceptions and expressions of moral principles. Therefore, future research should aim to investigate the role of cultural and linguistic factors in shaping the expression and interpretation of moral principles in OSS communities.



\vspace{-10pt}

\section{Conclusion}
This paper investigates moral principles and their relation with toxicity in OSS.
Due to the use of domain-specific words and jargon, the majority of toxicity detection tools do not correctly identify toxic language in OSS. Therefore, there is a need to augment the current methods with informative features, and to gain a comprehensive understanding of the triggers of toxicity and its underlying causes.
In this work, we leverage a toxicity-labeled dataset of GitHub to analyze the key concepts of five moral principles and identify exhibited patterns in the dataset. We also mapped the nature of toxicity to the observed types of moral principles in each thread. Our findings revealed that \textit{Sanctity/Degradation} and \textit{Care/Harm} were the most frequently observed moral principles, and were often associated with insulting threads.
Overall, our study represents an initial step towards a deeper understanding of the role of moral principles in shaping community dynamics in OSS, and we hope that it will inspire further research in this important area. Consistent with Miller et al.'s ~\cite{Miller2022} data distribution policy, we will share our annotated dataset upon request for research purposes.
\bibliographystyle{ACM-Reference-Format}
\bibliography{fse, aaai22,oss,preetha}

\end{document}